\documentclass[dvips]{article}
\usepackage{supertabular,lscape,epsfig}
\usepackage{amssymb}
\usepackage{amsmath}

\DeclareSymbolFont{ppa}{OT1}{ppl}{m}{it}
\DeclareMathSymbol{\vv}{\mathalpha}{ppa}{'166}

\begin{document}

\newcommand{\dd}{\,{\rm d}}
\newcommand{\ie}{{\it i.e.},\,}
\newcommand{\etal}{{\it et al.\ }}
\newcommand{\eg}{{\it e.g.},\,}
\newcommand{\cf}{{\it cf.\ }}
\newcommand{\vs}{{\it vs.\ }}
\newcommand{\zdot}{\makebox[0pt][l]{.}}
\newcommand{\up}[1]{\ifmmode^{\rm #1}\else$^{\rm #1}$\fi}
\newcommand{\dn}[1]{\ifmmode_{\rm #1}\else$_{\rm #1}$\fi}
\newcommand{\upd}{\up{d}}
\newcommand{\uph}{\up{h}}
\newcommand{\upm}{\up{m}}
\newcommand{\ups}{\up{s}}
\newcommand{\arcd}{\ifmmode^{\circ}\else$^{\circ}$\fi}
\newcommand{\arcm}{\ifmmode{'}\else$'$\fi}
\newcommand{\arcs}{\ifmmode{''}\else$''$\fi}
\newcommand{\MS}{{\rm M}\ifmmode_{\odot}\else$_{\odot}$\fi}
\newcommand{\RS}{{\rm R}\ifmmode_{\odot}\else$_{\odot}$\fi}
\newcommand{\LS}{{\rm L}\ifmmode_{\odot}\else$_{\odot}$\fi}

\newcommand{\Abstract}[2]{{\footnotesize\begin{center}ABSTRACT\end{center}
\vspace{1mm}\par#1\par
\noindent
{~}{\it #2}}}

\newcommand{\TabCap}[2]{\begin{center}\parbox[t]{#1}{\begin{center}
  \small {\spaceskip 2pt plus 1pt minus 1pt T a b l e}
  \refstepcounter{table}\thetable \\[2mm]
  \footnotesize #2 \end{center}}\end{center}}

\newcommand{\TableSep}[2]{\begin{table}[p]\vspace{#1}
\TabCap{#2}\end{table}}

\newcommand{\FigCap}[1]{\footnotesize\par\noindent Fig.\  %
  \refstepcounter{figure}\thefigure. #1\par}

\newcommand{\TableFont}{\footnotesize}
\newcommand{\TableFontIt}{\ttit}
\newcommand{\SetTableFont}[1]{\renewcommand{\TableFont}{#1}}

\newcommand{\MakeTable}[4]{\begin{table}[htb]\TabCap{#2}{#3}
  \begin{center} \TableFont \begin{tabular}{#1} #4 
  \end{tabular}\end{center}\end{table}}

\newcommand{\MakeTableSep}[4]{\begin{table}[p]\TabCap{#2}{#3}
  \begin{center} \TableFont \begin{tabular}{#1} #4 
  \end{tabular}\end{center}\end{table}}

\newenvironment{references}%
{
\footnotesize \frenchspacing
\renewcommand{\thesection}{}
\renewcommand{\in}{{\rm in }}
\renewcommand{\AA}{Astron.\ Astrophys.}
\newcommand{\AAS}{Astron.~Astrophys.~Suppl.~Ser.}
\newcommand{\ApJ}{Astrophys.\ J.}
\newcommand{\ApJS}{Astrophys.\ J.~Suppl.~Ser.}
\newcommand{\ApJL}{Astrophys.\ J.~Letters}
\newcommand{\AJ}{Astron.\ J.}
\newcommand{\IBVS}{IBVS}
\newcommand{\PASP}{P.A.S.P.}
\newcommand{\Acta}{Acta Astron.}
\newcommand{\MNRAS}{MNRAS}
\renewcommand{\and}{{\rm and }}
\section{{\rm REFERENCES}}
\sloppy \hyphenpenalty10000
\begin{list}{}{\leftmargin1cm\listparindent-1cm
\itemindent\listparindent\parsep0pt\itemsep0pt}}%
{\end{list}\vspace{2mm}}

\def\TYLDA{~}
\newlength{\DW}
\settowidth{\DW}{0}
\newcommand{\dw}{\hspace{\DW}}

\newcommand{\refitem}[5]{\item[]{#1} #2%
\def\REFARG{#3}\ifx\REFARG\TYLDA\else, {\it#3}\fi
\def\REFARG{#4}\ifx\REFARG\TYLDA\else, {\bf#4}\fi
\def\REFARG{#5}\ifx\REFARG\TYLDA\else, {#5}\fi.}

\newcommand{\Section}[1]{\section{#1}}
\newcommand{\Subsection}[1]{\subsection{#1}}
\newcommand{\Acknow}[1]{\par\vspace{5mm}{\bf Acknowledgements.} #1}
\pagestyle{myheadings}

\newfont{\bb}{ptmbi8t at 12pt}
\newcommand{\xrule}{\rule{0pt}{2.5ex}}
\newcommand{\xxrule}{\rule[-1.8ex]{0pt}{4.5ex}}
\def\thefootnote{\fnsymbol{footnote}}
\begin{center}
{\Large\bf The Optical Gravitational Lensing Experiment.
Optical Monitoring of the Gravitationally Lensed Quasar HE1104\,--1805
in 1997--2002\footnote{Based on  observations obtained with the 1.3~m
Warsaw telescope at the Las Campanas  Observatory of the Carnegie
Institution of Washington.}}

\vskip1cm
{\bf \L.~~W~y~r~z~y~k~o~w~s~k~i$^1$,
~~A.~~U~d~a~l~s~k~i$^1$,~~P.~L.~~S~c~h~e~c~h~t~e~r$^2$,\\
~~O.~~S~z~e~w~c~z~y~k$^1$,~~M.~~S~z~y~m~a~{\'n}~s~k~i$^1$, ~~M.~~K~u~b~i~a~k$^1$,\\ 
G.~~P~i~e~t~r~z~y~\'n~s~k~i$^{1}$,~~ I.~~S~o~s~z~y~\'n~s~k~i~~ and ~~ K.~~\.Z~e~b~r~u~\'n$^1$}
\vskip3mm
{$^1$Warsaw University Observatory, Al.~Ujazdowskie~4, 00-478~Warszawa,
Poland\\
e-mail: (wyrzykow,udalski,szewczyk,msz,mk,pietrzyn,soszynsk,zebrun)@astrouw.edu.pl\\
$^2$ Massachusetts Institute of Technology, 77 Massachusetts Avenue,
Cambridge, MA 02139-4307, USA\\
e-mail: schech@achernar.mit.edu}
\end{center}

\Abstract{We present results of the long term monitoring of the 
gravitationally lensed quasar HE1104\,--1805. The photometric data were 
collected between August 1997 and January 2002 as a subproject of the OGLE 
survey. 

We determine the time delay in the light curves of images A and B of 
HE1104\,--1805 to be equal to ${157\pm21}$ days with the variability in the 
image B light curve leading variability of the image A. The result is in 
excellent agreement with the earlier determination by Ofek and Maoz. 

OGLE photometry of HE1104\,--1805 is available to the astronomical community 
from the OGLE Internet archive.}{}

\Section{Introduction}
Time delay between variability pattern in the images of gravitationally lensed 
quasars has been measured in several cases (\eg Schechter \etal 1997, Kundi\'c 
\etal 1997). Determination of time delay has important consequences, as it 
makes it possible to measure the Hubble constant at long cosmological 
distances. 

The gravitationally lensed quasar HE1104\,--1805 with relatively large 
separation between two images A and B (about 3\arcs) was discovered by 
Wisotzki \etal (1993). The object has been included to the list of regularly 
monitored objects by the Optical Gravitational Lensing Experiment (OGLE) at 
the beginning of the second phase of the project -- OGLE-II (Udalski, 
Kubiak and Szyma{\'n}ski 1997) in August 1997. The main goal of this 
sub-project of the OGLE survey was a determination of the time delay of 
HE1104\,--1805. 

The quasar was regularly observed up to the end of the 1999/2000 observing 
season (August~2000) with frequency of one observation per 5--7 days (except 
for three month period, August--October, each year when the quasar was not 
visible from the Earth). In this way a unique dataset on the long term 
photometric behavior of HE1104\,--1805 was collected. Unfortunately, the main 
goal of the monitoring -- determination of the time delay -- has not been 
accomplished (Schechter \etal 2003). It turned out that both images of 
HE1104\,--1805 changed the brightness in a rather uncorrelated way making 
determination of time delay practically impossible. 

On the other hand, the large variability of image A of HE1104\,--1805 on a 
time scale of weeks was most likely caused by microlensing. Variability of a 
gravitationally lensed quasar is believed to consist of at least two 
components: intrinsic variability of the lensed quasar and microlensing 
variability caused by lensing galaxy. Thus, surprisingly, the dataset 
collected by OGLE-II provided unique observational material for studying 
microlensing in gravitational lenses and contributed to better understanding 
of this phenomenon (Schechter \etal 2003). 

In November~2000 the OGLE-II phase ended and observations were suspended. 
After a half year break the OGLE project resumed observations after a 
significant upgrade of observing capabilities (OGLE-III). HE1104\,--1805 was 
still on the list of targets, and the object was regularly observed up to 
January 2002. Unfortunately, the OGLE-II data pipeline providing photometry of 
the quasar in the real time was not adjusted to the OGLE-III phase at that 
time. Therefore, the collected data were stored and photometry of 
HE1104\,--1805 from 2001--2002 was not available at the time of our first 
analysis (Schechter \etal 2003). 

In the meanwhile Ofek and Maoz (2003) announced a determination of the time 
delay of HE1104\,--1805 of ${161\pm7}$ days. They used the OGLE-II photometry 
combined with their own observations in the {\it R}-band collected from 1999 
to 2002. The data presented by Ofek and Maoz (2003) indicated that at the time 
when the OGLE-II suspended observations for the hardware transition to the 
OGLE-III phase, the B component of HE1104\,--1805 faded by almost 0.3~mag and 
the pattern was then followed by the A component providing good opportunity 
for time delay measurement. 

The results of Ofek and Maoz (2003) stimulated us to reduce the data collected 
during OGLE-III phase (2001--2002) and a few additional observations collected 
at the end of the OGLE-II phase and not presented in Schechter \etal (2003). 
We also attempted to independently derive the time delay based solely on the 
OGLE dataset of 1997--2002 which is more uniform (only {\it V}-band 
measurements) and accurate than that of Ofek and Maoz (2003). We present our 
results in this paper. 

\Section{Observations}
Observations of HE1104\,--1805 were made between August 1997 and January 2002 
as a subproject of the OGLE-II and OGLE-III surveys. All data were collected 
with the 1.3-m Warsaw telescope at the Las Campanas Observatory in Chile, 
operated by the Carnegie Institution of Washington. 

\begin{figure}[htb]
\centerline{\includegraphics[width=13cm]{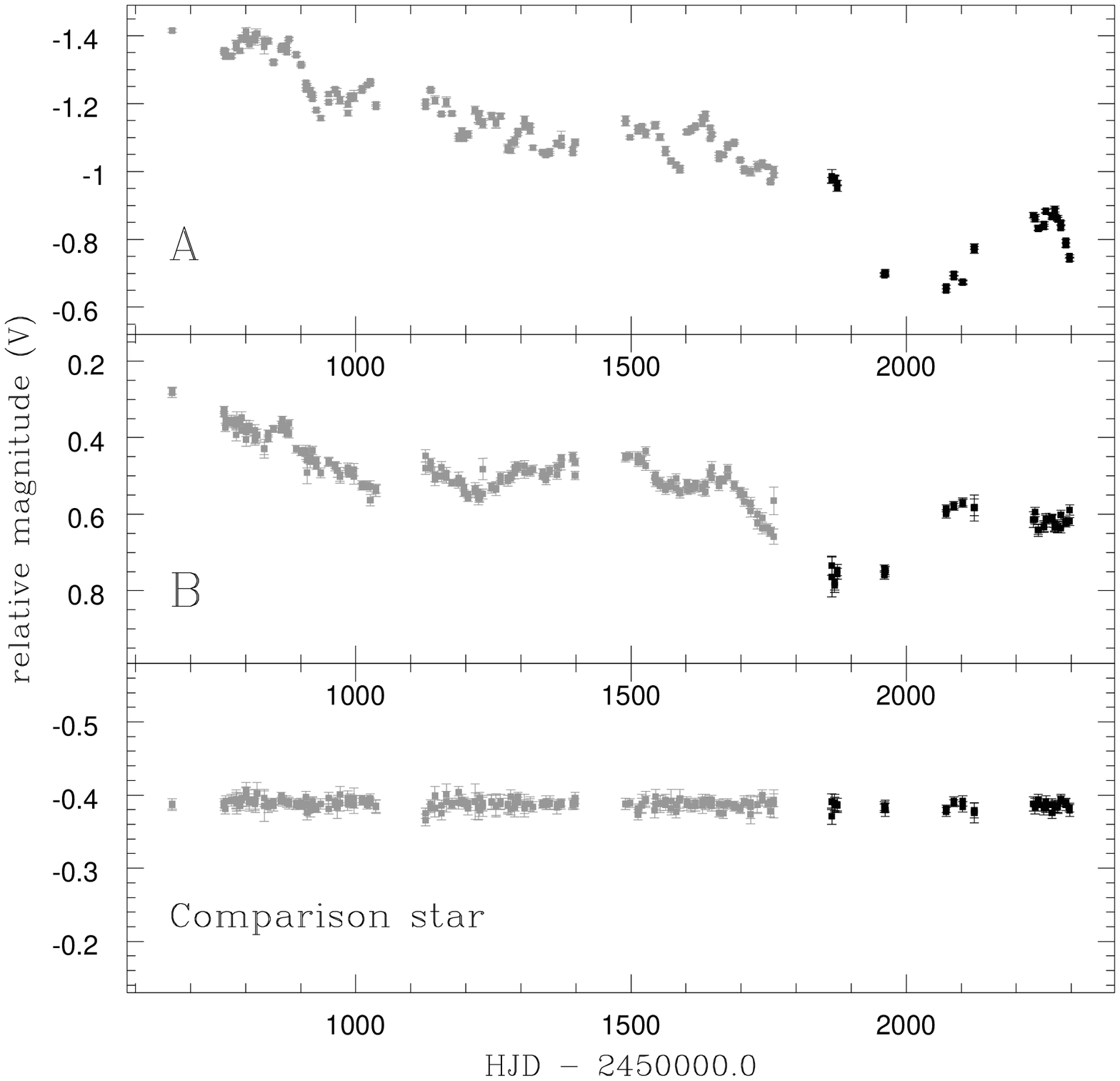}}
\FigCap{Light curves of image A (upper panel) and B (middle panel) 
of the gravitationally lensed quasar HE1104\,--1805 collected during the OGLE 
monitoring. In the bottom panel brightness of the nearby comparison 
stars, CA, is shown. Grey points indicate data presented in Schechter \etal 
(2003).} 
\end{figure}
Observations collected up to August 2000 were described in Schechter \etal 
(2003). After the conjunction with the Sun the quasar was observed a couple of 
times in November 2000. Unfortunately, because of autoguider failure at that 
time in each case five 2 minutes exposures were obtained instead of the 
standard 10 minute one. The 2 minute images were then aligned, stacked  and 
added before the photometry was derived. Additional observation of 
HE1104\,--1805 was obtained in February 2001. This time two standard 10 minute 
exposures were taken. All these observations were made with the OGLE-II setup 
described in Schechter \etal (2003). 

From June 2001 to January 2002, HE1104\,--1805 was observed with OGLE-III 
setup, \ie with the ${8192\times 8192}$ pixel eight chip mosaic camera 
(Udalski \etal 2002) The frequency of observations was similar as during the 
OGLE-II coverage. Each observation consisted of two 10 minute exposures and 
the field was shifted a few arcsec between them. 

The new images of HE1104\,--1805 were reduced in the identical manner as the 
earlier ones (Schechter \etal 2003). Photometry was derived with the {\sc 
DoPhot} photometry program (Schechter, Mateo and Saha 1993). The same 
comparison stars were used for consistency. The finding chart of HE1104\,--1805 
and positions of comparison stars can be found in Fig.~1 in Schechter \etal 
(2003). 

Our Fig.~1 presents the entire light curve of HE1104\,--1805 collected during 
OGLE observations. Individual measurements are listed in Table~1 and are also 
available to the astronomical community in digital form from the OGLE archive: 

\begin{center}
{\it http://ogle.astrouw.edu.pl}\\
{\it ftp://sirius.astrouw.edu.pl/ogle/ogle2/HE1104/}\\
\end{center}
or its US mirror
\begin{center}
{\it http://bulge.princeton.edu/\~{}ogle}\\
{\it ftp://bulge.princeton.edu/ogle/ogle2/HE1104/}\\
\end{center}

\renewcommand{\arraystretch}{0.87}
\renewcommand{\TableFont}{\scriptsize}
\MakeTableSep{ccccccc}{12.5cm}{Photometry of HE1104\,--1805}
{\hline\noalign{\vskip4pt}
HJD --
&\multicolumn{2}{c}{$\rm Q_A$}&\multicolumn{2}{c}{$\rm Q_B$}&\multicolumn{2}{c}{CA}\\
2450000    &  $V$    & $\sigma_V$  & $V$   & $\sigma_V$   &  $V$   & $\sigma_V$ \\   
\noalign{\vskip4pt}\hline\noalign{\vskip4pt}
 666.46490 & $-1.415$ & 0.005 & 0.283 & 0.012 & $-0.388$ & 0.007\\
 666.47390 & $-1.415$ & 0.005 & 0.277 & 0.009 & $-0.387$ & 0.008\\
 760.84802 & $-1.357$ & 0.003 & 0.338 & 0.007 & $-0.389$ & 0.006\\
 760.85665 & $-1.351$ & 0.003 & 0.327 & 0.009 & $-0.385$ & 0.006\\
 762.85473 & $-1.338$ & 0.004 & 0.373 & 0.011 & $-0.381$ & 0.007\\
 763.85505 & $-1.354$ & 0.006 & 0.357 & 0.010 & $-0.392$ & 0.007\\
 773.84775 & $-1.338$ & 0.005 & 0.362 & 0.011 & $-0.393$ & 0.008\\
 773.85593 & $-1.341$ & 0.005 & 0.354 & 0.012 & $-0.392$ & 0.007\\
 782.84199 & $-1.372$ & 0.013 & 0.392 & 0.016 & $-0.394$ & 0.013\\
 782.85017 & $-1.375$ & 0.012 & 0.365 & 0.015 & $-0.393$ & 0.011\\
 783.84383 & $-1.363$ & 0.015 & 0.352 & 0.019 & $-0.389$ & 0.015\\
 783.85217 & $-1.375$ & 0.012 & 0.362 & 0.014 & $-0.390$ & 0.012\\
 789.85059 & $-1.356$ & 0.007 & 0.361 & 0.008 & $-0.385$ & 0.006\\
 792.81232 & $-1.389$ & 0.006 & 0.380 & 0.009 & $-0.396$ & 0.007\\
 792.82112 & $-1.395$ & 0.007 & 0.348 & 0.015 & $-0.394$ & 0.009\\
 800.84329 & $-1.405$ & 0.005 & 0.385 & 0.012 & $-0.403$ & 0.008\\
 800.85268 & $-1.412$ & 0.012 & 0.370 & 0.016 & $-0.397$ & 0.013\\
 800.86372 & $-1.388$ & 0.006 & 0.405 & 0.018 & $-0.407$ & 0.010\\
 806.85209 & $-1.375$ & 0.012 & 0.380 & 0.013 & $-0.390$ & 0.012\\
 806.86044 & $-1.387$ & 0.013 & 0.369 & 0.013 & $-0.394$ & 0.013\\
 816.86178 & $-1.386$ & 0.013 & 0.404 & 0.016 & $-0.393$ & 0.014\\
 816.87297 & $-1.402$ & 0.009 & 0.381 & 0.015 & $-0.388$ & 0.007\\
 817.84615 & $-1.403$ & 0.018 & 0.396 & 0.017 & $-0.392$ & 0.016\\
 817.85528 & $-1.395$ & 0.015 & 0.402 & 0.014 & $-0.394$ & 0.013\\
 820.86843 & $-1.407$ & 0.014 & 0.393 & 0.016 & $-0.402$ & 0.015\\
 833.80560 & $-1.386$ & 0.012 & 0.428 & 0.016 & $-0.396$ & 0.012\\
 833.81395 & $-1.368$ & 0.022 & 0.429 & 0.024 & $-0.385$ & 0.021\\
 840.79155 & $-1.386$ & 0.005 & 0.402 & 0.009 & $-0.390$ & 0.006\\
 840.79989 & $-1.383$ & 0.008 & 0.389 & 0.009 & $-0.384$ & 0.008\\
 850.81350 & $-1.324$ & 0.005 & 0.378 & 0.008 & $-0.391$ & 0.005\\
 850.82185 & $-1.320$ & 0.006 & 0.376 & 0.008 & $-0.385$ & 0.007\\
 864.65372 & $-1.359$ & 0.004 & 0.381 & 0.006 & $-0.392$ & 0.005\\
 864.66251 & $-1.366$ & 0.004 & 0.365 & 0.008 & $-0.399$ & 0.005\\
 866.77954 & $-1.370$ & 0.005 & 0.370 & 0.007 & $-0.392$ & 0.007\\
 866.78846 & $-1.361$ & 0.004 & 0.353 & 0.009 & $-0.394$ & 0.006\\
 874.76657 & $-1.351$ & 0.006 & 0.389 & 0.009 & $-0.390$ & 0.007\\
 874.77512 & $-1.371$ & 0.006 & 0.376 & 0.010 & $-0.391$ & 0.008\\
 878.66866 & $-1.391$ & 0.004 & 0.389 & 0.011 & $-0.391$ & 0.006\\
 878.67720 & $-1.390$ & 0.004 & 0.363 & 0.010 & $-0.388$ & 0.007\\
 891.74182 & $-1.346$ & 0.005 & 0.430 & 0.010 & $-0.386$ & 0.006\\
 891.75040 & $-1.343$ & 0.006 & 0.430 & 0.010 & $-0.389$ & 0.006\\
 900.64439 & $-1.312$ & 0.004 & 0.439 & 0.008 & $-0.386$ & 0.005\\
 900.65275 & $-1.317$ & 0.004 & 0.433 & 0.007 & $-0.390$ & 0.006\\
 909.58239 & $-1.242$ & 0.005 & 0.441 & 0.016 & $-0.388$ & 0.008\\
 909.59072 & $-1.262$ & 0.005 & 0.435 & 0.015 & $-0.397$ & 0.008\\
 911.77374 & $-1.247$ & 0.007 & 0.459 & 0.020 & $-0.376$ & 0.010\\
 911.78210 & $-1.253$ & 0.007 & 0.491 & 0.029 & $-0.386$ & 0.012\\
 917.68358 & $-1.227$ & 0.006 & 0.451 & 0.019 & $-0.379$ & 0.009\\
 917.69199 & $-1.240$ & 0.006 & 0.461 & 0.018 & $-0.387$ & 0.009\\
 921.64575 & $-1.214$ & 0.004 & 0.460 & 0.010 & $-0.391$ & 0.006\\
 921.65410 & $-1.228$ & 0.004 & 0.433 & 0.011 & $-0.390$ & 0.008\\
 928.62998 & $-1.181$ & 0.004 & 0.459 & 0.009 & $-0.382$ & 0.006\\
 928.63834 & $-1.180$ & 0.005 & 0.474 & 0.009 & $-0.383$ & 0.006\\
 936.57571 & $-1.157$ & 0.006 & 0.492 & 0.013 & $-0.388$ & 0.007\\
 950.57474 & $-1.204$ & 0.004 & 0.464 & 0.011 & $-0.381$ & 0.007\\
 950.58309 & $-1.228$ & 0.006 & 0.462 & 0.011 & $-0.396$ & 0.008\\
 962.52107 & $-1.243$ & 0.005 & 0.471 & 0.009 & $-0.391$ & 0.006\\
 962.52943 & $-1.240$ & 0.006 & 0.476 & 0.011 & $-0.387$ & 0.008\\
 966.54958 & $-1.229$ & 0.016 & 0.486 & 0.018 & $-0.384$ & 0.016\\
\hline}
\setcounter{table}{0}
\MakeTableSep{ccccccc}{12.5cm}{Continued}
{\hline\noalign{\vskip4pt}
HJD --
&\multicolumn{2}{c}{$\rm Q_A$}&\multicolumn{2}{c}{$\rm Q_B$}&\multicolumn{2}{c}{CA}\\
2450000    &  $V$    & $\sigma_V$  & $V$   & $\sigma_V$   &  $V$   & $\sigma_V$ \\   
\noalign{\vskip4pt}\hline\noalign{\vskip4pt} 
~~971.46710 & $-1.211$ & 0.010 & 0.496 & 0.018 & $-0.401$ & 0.011\\
~~971.47701 & $-1.209$ & 0.011 & 0.502 & 0.017 & $-0.389$ & 0.011\\
~~971.47701 & $-1.209$ & 0.011 & 0.502 & 0.017 & $-0.389$ & 0.011\\
~~985.49468 & $-1.198$ & 0.009 & 0.479 & 0.013 & $-0.396$ & 0.010\\
~~985.50301 & $-1.172$ & 0.009 & 0.491 & 0.011 & $-0.388$ & 0.009\\
~~990.48608 & $-1.214$ & 0.006 & 0.491 & 0.010 & $-0.396$ & 0.006\\
~~990.49443 & $-1.223$ & 0.005 & 0.493 & 0.010 & $-0.390$ & 0.007\\
~~996.46974 & $-1.218$ & 0.012 & 0.484 & 0.016 & $-0.387$ & 0.012\\
~~996.47811 & $-1.224$ & 0.015 & 0.501 & 0.020 & $-0.396$ & 0.014\\
1011.49065 & $-1.245$ & 0.005 & 0.528 & 0.009 & $-0.395$ & 0.005\\
1011.49901 & $-1.238$ & 0.004 & 0.522 & 0.010 & $-0.391$ & 0.005\\
1020.47797 & $-1.256$ & 0.004 & 0.528 & 0.009 & $-0.393$ & 0.007\\
1020.48633 & $-1.254$ & 0.004 & 0.523 & 0.009 & $-0.387$ & 0.006\\
1026.47489 & $-1.265$ & 0.004 & 0.527 & 0.014 & $-0.395$ & 0.007\\
1026.48324 & $-1.259$ & 0.005 & 0.563 & 0.015 & $-0.389$ & 0.007\\
1036.47650 & $-1.191$ & 0.006 & 0.533 & 0.012 & $-0.385$ & 0.009\\
1036.46814 & $-1.198$ & 0.006 & 0.539 & 0.015 & $-0.384$ & 0.009\\
1126.83991 & $-1.206$ & 0.005 & 0.480 & 0.016 & $-0.375$ & 0.007\\
1126.84826 & $-1.191$ & 0.005 & 0.448 & 0.016 & $-0.366$ & 0.008\\
1135.83662 & $-1.242$ & 0.004 & 0.479 & 0.011 & $-0.388$ & 0.006\\
1135.84498 & $-1.238$ & 0.005 & 0.465 & 0.011 & $-0.382$ & 0.006\\
1143.84217 & $-1.208$ & 0.010 & 0.509 & 0.016 & $-0.399$ & 0.013\\
1143.85053 & $-1.211$ & 0.012 & 0.496 & 0.019 & $-0.382$ & 0.012\\
1155.83927 & $-1.169$ & 0.005 & 0.477 & 0.015 & $-0.386$ & 0.007\\
1155.84763 & $-1.170$ & 0.004 & 0.501 & 0.014 & $-0.375$ & 0.009\\
1164.81567 & $-1.203$ & 0.013 & 0.504 & 0.015 & $-0.401$ & 0.014\\
1164.82404 & $-1.206$ & 0.014 & 0.493 & 0.015 & $-0.391$ & 0.014\\
1174.79099 & $-1.170$ & 0.006 & 0.518 & 0.008 & $-0.389$ & 0.007\\
1174.79932 & $-1.172$ & 0.005 & 0.518 & 0.009 & $-0.387$ & 0.006\\
1186.81180 & $-1.097$ & 0.007 & 0.519 & 0.013 & $-0.404$ & 0.008\\
1186.82014 & $-1.105$ & 0.008 & 0.506 & 0.016 & $-0.390$ & 0.009\\
1192.75985 &  --      & --    & 0.523 & 0.011 & $-0.389$ & 0.008\\
1192.76819 & $-1.121$ & 0.008 & 0.515 & 0.010 & $-0.393$ & 0.008\\
1196.78326 & $-1.096$ & 0.005 & 0.545 & 0.009 & $-0.391$ & 0.006\\
1196.79160 & $-1.110$ & 0.005 & 0.530 & 0.009 & $-0.388$ & 0.007\\
1203.78371 & $-1.106$ & 0.007 & 0.558 & 0.009 & $-0.389$ & 0.006\\
1203.79205 & $-1.112$ & 0.008 & 0.549 & 0.009 & $-0.382$ & 0.009\\
1216.77722 & $-1.182$ & 0.011 & 0.540 & 0.015 & $-0.391$ & 0.012\\
1216.78557 & $-1.180$ & 0.012 & 0.534 & 0.016 & $-0.391$ & 0.024\\
1222.76902 & $-1.158$ & 0.015 & 0.557 & 0.018 & $-0.383$ & 0.016\\
1222.77738 & $-1.162$ & 0.017 & 0.541 & 0.017 & $-0.384$ & 0.018\\
1224.75571 & $-1.147$ & 0.008 & 0.554 & 0.011 & $-0.379$ & 0.008\\
1224.76405 & $-1.171$ & 0.007 & 0.559 & 0.010 & $-0.396$ & 0.008\\
1230.78547 & $-1.139$ & 0.011 & 0.482 & 0.028 & $-0.384$ & 0.013\\
1230.79383 & $-1.147$ & 0.010 & 0.547 & 0.013 & $-0.389$ & 0.011\\
1246.70067 & $-1.165$ & 0.009 & 0.532 & 0.013 & $-0.391$ & 0.010\\
1246.70901 & $-1.161$ & 0.009 & 0.528 & 0.010 & $-0.390$ & 0.010\\
1254.74656 & $-1.145$ & 0.016 & 0.533 & 0.018 & $-0.386$ & 0.014\\
1254.75492 & $-1.141$ & 0.014 & 0.535 & 0.016 & $-0.386$ & 0.016\\
1262.73071 & $-1.164$ & 0.008 & 0.500 & 0.011 & $-0.391$ & 0.010\\
1262.73907 & $-1.162$ & 0.007 & 0.514 & 0.009 & $-0.388$ & 0.008\\
1276.59142 & $-1.070$ & 0.011 & 0.504 & 0.014 & $-0.388$ & 0.012\\
1276.59975 & $-1.065$ & 0.010 & 0.513 & 0.013 & $-0.385$ & 0.011\\
1281.66500 & $-1.085$ & 0.008 & 0.497 & 0.011 & $-0.398$ & 0.009\\
1281.67333 & $-1.061$ & 0.007 & 0.507 & 0.010 & $-0.381$ & 0.008\\
1288.62525 & $-1.084$ & 0.012 & 0.492 & 0.015 & $-0.383$ & 0.013\\
1288.63359 & $-1.093$ & 0.009 & 0.477 & 0.013 & $-0.389$ & 0.011\\
1294.62164 & $-1.110$ & 0.005 & 0.479 & 0.012 & $-0.398$ & 0.006\\
1294.62998 & $-1.119$ & 0.005 & 0.474 & 0.012 & $-0.390$ & 0.007\\
\hline}
\setcounter{table}{0}
\MakeTableSep{ccccccc}{12.5cm}{Continued}
{\hline\noalign{\vskip4pt}
HJD --
&\multicolumn{2}{c}{$\rm Q_A$}&\multicolumn{2}{c}{$\rm Q_B$}&\multicolumn{2}{c}{CA}\\
2450000    &  $V$    & $\sigma_V$  & $V$   & $\sigma_V$   &  $V$   & $\sigma_V$ \\   
\noalign{\vskip4pt}\hline\noalign{\vskip4pt}
1306.60288 & $-1.133$ & 0.011 & 0.491 & 0.015 & $-0.388$ & 0.013\\
1306.61120 & $-1.152$ & 0.011 & 0.473 & 0.013 & $-0.382$ & 0.012\\
1316.63431 & $-1.120$ & 0.010 & 0.490 & 0.011 & $-0.388$ & 0.010\\
1316.64266 & $-1.135$ & 0.005 & 0.487 & 0.010 & $-0.388$ & 0.007\\
1321.63638 & $-1.070$ & 0.006 & 0.482 & 0.010 & $-0.384$ & 0.007\\
1339.49695 & $-1.057$ & 0.004 & 0.501 & 0.008 & $-0.387$ & 0.005\\
1339.50534 & $-1.056$ & 0.003 & 0.490 & 0.009 & $-0.390$ & 0.004\\
1345.49811 & $-1.049$ & 0.006 & 0.501 & 0.010 & $-0.391$ & 0.008\\
1345.50645 & $-1.053$ & 0.005 & 0.511 & 0.012 & $-0.387$ & 0.007\\
1352.45965 & $-1.053$ & 0.006 & 0.489 & 0.013 & $-0.390$ & 0.009\\
1352.46797 & $-1.060$ & 0.006 & 0.485 & 0.012 & $-0.388$ & 0.008\\
1365.47141 & $-1.081$ & 0.008 & 0.497 & 0.011 & $-0.383$ & 0.009\\
1365.47976 & $-1.084$ & 0.007 & 0.476 & 0.011 & $-0.387$ & 0.007\\
1373.47231 & $-1.076$ & 0.004 & 0.452 & 0.008 & $-0.390$ & 0.005\\
1373.48063 & $-1.099$ & 0.020 & 0.470 & 0.014 & $-0.387$ & 0.006\\
1394.48817 & $-1.065$ & 0.006 & 0.448 & 0.010 & $-0.387$ & 0.007\\
1394.49649 & $-1.054$ & 0.005 & 0.454 & 0.010 & $-0.389$ & 0.007\\
1398.46873 & $-1.088$ & 0.007 & 0.499 & 0.011 & $-0.396$ & 0.008\\
1398.47706 & $-1.083$ & 0.007 & 0.464 & 0.010 & $-0.388$ & 0.007\\
1489.84879 & $-1.143$ & 0.008 & 0.449 & 0.010 & $-0.388$ & 0.008\\
1489.85714 & $-1.156$ & 0.008 & 0.452 & 0.012 & $-0.388$ & 0.008\\
1497.85002 & $-1.100$ & 0.004 & 0.448 & 0.011 & $-0.389$ & 0.006\\
1512.82977 & $-1.129$ & 0.005 & 0.448 & 0.011 & $-0.373$ & 0.007\\
1512.83812 & $-1.118$ & 0.005 & 0.464 & 0.011 & $-0.382$ & 0.006\\
1519.82916 & $-1.133$ & 0.006 & 0.455 & 0.009 & $-0.388$ & 0.013\\
1519.83753 & $-1.126$ & 0.009 & 0.460 & 0.011 & $-0.386$ & 0.010\\
1526.83854 & $-1.116$ & 0.006 & 0.435 & 0.011 & $-0.392$ & 0.007\\
1526.84689 & $-1.109$ & 0.011 & 0.473 & 0.013 & $-0.394$ & 0.011\\
1542.85348 & $-1.134$ & 0.010 & 0.504 & 0.013 & $-0.379$ & 0.010\\
1544.81970 & $-1.135$ & 0.006 & 0.517 & 0.011 & $-0.388$ & 0.007\\
1544.82803 & $-1.138$ & 0.009 & 0.496 & 0.011 & $-0.398$ & 0.010\\
1552.83863 & $-1.101$ & 0.010 & 0.526 & 0.014 & $-0.388$ & 0.012\\
1552.84703 & $-1.103$ & 0.007 & 0.514 & 0.011 & $-0.389$ & 0.010\\
1562.84324 & $-1.065$ & 0.008 & 0.536 & 0.013 & $-0.387$ & 0.008\\
1562.85157 & $-1.054$ & 0.007 & 0.526 & 0.011 & $-0.393$ & 0.008\\
1572.81787 & $-1.032$ & 0.005 & 0.516 & 0.014 & $-0.392$ & 0.016\\
1572.82623 & $-1.030$ & 0.007 & 0.527 & 0.014 & $-0.382$ & 0.009\\
1581.77920 & $-1.018$ & 0.008 & 0.531 & 0.013 & $-0.387$ & 0.009\\
1581.78754 & $-1.020$ & 0.006 & 0.507 & 0.013 & $-0.377$ & 0.008\\
1588.77023 & $-1.010$ & 0.007 & 0.545 & 0.012 & $-0.395$ & 0.013\\
1588.78201 & $-1.003$ & 0.007 & 0.541 & 0.010 & $-0.393$ & 0.015\\
1601.78041 & $-1.118$ & 0.006 & 0.538 & 0.010 & $-0.387$ & 0.008\\
1601.78873 & $-1.115$ & 0.006 & 0.518 & 0.010 & $-0.391$ & 0.007\\
1608.76978 & $-1.119$ & 0.005 & 0.528 & 0.009 & $-0.385$ & 0.006\\
1608.77813 & $-1.126$ & 0.006 & 0.535 & 0.011 & $-0.385$ & 0.008\\
1616.75219 & $-1.129$ & 0.004 & 0.519 & 0.008 & $-0.390$ & 0.006\\
1616.76053 & $-1.135$ & 0.003 & 0.531 & 0.011 & $-0.385$ & 0.007\\
1629.67381 & $-1.142$ & 0.010 & 0.533 & 0.017 & $-0.386$ & 0.011\\
1629.68215 & $-1.160$ & 0.007 & 0.527 & 0.012 & $-0.392$ & 0.009\\
1634.66890 & $-1.170$ & 0.007 & 0.520 & 0.011 & $-0.393$ & 0.008\\
1634.67724 & $-1.156$ & 0.007 & 0.541 & 0.012 & $-0.389$ & 0.009\\
1643.64098 & $-1.096$ & 0.003 & 0.495 & 0.009 & $-0.387$ & 0.005\\
1643.64934 & $-1.130$ & 0.005 & 0.493 & 0.011 & $-0.394$ & 0.006\\
1646.58363 & $-1.108$ & 0.005 & 0.506 & 0.011 & $-0.385$ & 0.007\\
1646.59197 & $-1.110$ & 0.006 & 0.477 & 0.015 & $-0.393$ & 0.007\\
1659.62581 & $-1.037$ & 0.004 & 0.526 & 0.012 & $-0.376$ & 0.006\\
1659.63416 & $-1.052$ & 0.004 & 0.511 & 0.010 & $-0.388$ & 0.006\\
1667.59933 & $-1.049$ & 0.005 & 0.512 & 0.010 & $-0.375$ & 0.007\\
1667.60776 & $-1.049$ & 0.006 & 0.510 & 0.011 & $-0.387$ & 0.008\\
\hline}
\renewcommand{\arraystretch}{0.85}
\setcounter{table}{0}
\MakeTableSep{ccccccc}{12.5cm}{Concluded}
{\hline\noalign{\vskip4pt}
HJD --
&\multicolumn{2}{c}{$\rm Q_A$}&\multicolumn{2}{c}{$\rm Q_B$}&\multicolumn{2}{c}{CA}\\
2450000    &  $V$    & $\sigma_V$  & $V$   & $\sigma_V$   &  $V$   & $\sigma_V$ \\   
\noalign{\vskip4pt}\hline\noalign{\vskip4pt}
1675.56667 & $-1.070$ & 0.004 & 0.481 & 0.011 & $-0.388$ & 0.007\\
1675.57503 & $-1.080$ & 0.004 & 0.501 & 0.012 & $-0.389$ & 0.006\\
1687.53458 & $-1.088$ & 0.004 & 0.522 & 0.009 & $-0.387$ & 0.005\\
1687.54291 & $-1.082$ & 0.003 & 0.529 & 0.011 & $-0.383$ & 0.005\\
1698.51723 & $-1.033$ & 0.005 & 0.538 & 0.008 & $-0.383$ & 0.007\\
1698.52560 & $-1.035$ & 0.004 & 0.549 & 0.008 & $-0.381$ & 0.006\\
1705.53958 & $-1.001$ & 0.004 & 0.567 & 0.012 & $-0.389$ & 0.006\\
1705.54792 & $-1.009$ & 0.004 & 0.548 & 0.014 & $-0.392$ & 0.007\\
1717.46476 & $-1.000$ & 0.008 & 0.574 & 0.011 & $-0.390$ & 0.008\\
1717.48314 & $-1.000$ & 0.014 & 0.571 & 0.015 & $-0.384$ & 0.016\\
1717.49147 & $-0.999$ & 0.011 & 0.592 & 0.015 & $-0.374$ & 0.013\\
1729.48827 & $-1.018$ & 0.012 & 0.600 & 0.014 & $-0.387$ & 0.012\\
1729.49661 & $-1.010$ & 0.011 & 0.623 & 0.015 & $-0.387$ & 0.011\\
1738.48134 & $-1.026$ & 0.005 & 0.610 & 0.014 & $-0.400$ & 0.007\\
1738.48970 & $-1.018$ & 0.005 & 0.636 & 0.015 & $-0.393$ & 0.008\\
1748.48952 & $-1.013$ & 0.004 & 0.636 & 0.011 & $-0.388$ & 0.005\\
1753.47627 & $-0.972$ & 0.005 & 0.645 & 0.011 & $-0.391$ & 0.006\\
1753.48460 & $-0.969$ & 0.006 & 0.647 & 0.011 & $-0.378$ & 0.008\\
1759.45499 & $-1.005$ & 0.011 & 0.565 & 0.036 & $-0.388$ & 0.019\\
1759.46335 & $-0.989$ & 0.007 & 0.658 & 0.020 & $-0.393$ & 0.009\\
1864.81454 & $-0.986$ & 0.019 & 0.764 & 0.052 & $-0.391$ & 0.011\\
1864.82704 & $-0.974$ & 0.009 & 0.734 & 0.025 & $-0.371$ & 0.011\\
1869.83225 & $-0.980$ & 0.007 & 0.786 & 0.018 & $-0.387$ & 0.010\\
1869.84549 & $-0.981$ & 0.008 & 0.778 & 0.020 & $-0.390$ & 0.011\\
1874.82956 & $-0.965$ & 0.008 & 0.756 & 0.014 & $-0.386$ & 0.010\\
1874.84185 & $-0.951$ & 0.009 & 0.746 & 0.015 & $-0.387$ & 0.009\\
1960.62892 & $-0.694$ & 0.005 & 0.742 & 0.010 & $-0.385$ & 0.006\\
1960.63726 & $-0.697$ & 0.005 & 0.759 & 0.011 & $-0.382$ & 0.006\\
1961.66697 & $-0.703$ & 0.008 & 0.750 & 0.013 & $-0.385$ & 0.008\\
1961.67531 & $-0.698$ & 0.009 & 0.747 & 0.011 & $-0.380$ & 0.009\\
2072.49533 & $-0.649$ & 0.005 & 0.599 & 0.011 & $-0.377$ & 0.006\\
2072.50381 & $-0.660$ & 0.005 & 0.585 & 0.009 & $-0.381$ & 0.005\\
2086.47804 & $-0.699$ & 0.005 & 0.576 & 0.010 & $-0.389$ & 0.005\\
2086.48652 & $-0.688$ & 0.004 & 0.581 & 0.009 & $-0.392$ & 0.006\\
2102.50286 & $-0.672$ & 0.006 & 0.572 & 0.010 & $-0.392$ & 0.007\\
2102.51139 & $-0.675$ & 0.006 & 0.568 & 0.010 & $-0.384$ & 0.007\\
2123.47288 & $-0.778$ & 0.008 & 0.582 & 0.022 & $-0.379$ & 0.010\\
2123.48140 & $-0.769$ & 0.010 & 0.584 & 0.034 & $-0.376$ & 0.014\\
2230.85928 & $-0.871$ & 0.008 & 0.614 & 0.021 & $-0.388$ & 0.010\\
2233.84468 & $-0.860$ & 0.008 & 0.594 & 0.012 & $-0.383$ & 0.009\\
2233.85322 & $-0.866$ & 0.008 & 0.613 & 0.013 & $-0.387$ & 0.009\\
2239.84689 & $-0.831$ & 0.006 & 0.640 & 0.013 & $-0.394$ & 0.007\\
2239.85545 & $-0.834$ & 0.007 & 0.642 & 0.016 & $-0.387$ & 0.009\\
2249.84233 & $-0.845$ & 0.007 & 0.631 & 0.014 & $-0.386$ & 0.008\\
2249.85085 & $-0.837$ & 0.007 & 0.634 & 0.014 & $-0.382$ & 0.009\\
2253.85220 & $-0.884$ & 0.005 & 0.612 & 0.010 & $-0.383$ & 0.007\\
2253.86073 & $-0.882$ & 0.006 & 0.616 & 0.018 & $-0.391$ & 0.008\\
2264.83957 & $-0.871$ & 0.007 & 0.610 & 0.010 & $-0.376$ & 0.008\\
2264.84811 & $-0.865$ & 0.006 & 0.614 & 0.011 & $-0.387$ & 0.008\\
2269.84004 & $-0.882$ & 0.007 & 0.624 & 0.010 & $-0.386$ & 0.007\\
2269.84857 & $-0.883$ & 0.009 & 0.634 & 0.012 & $-0.384$ & 0.009\\
2269.86228 & $-0.889$ & 0.009 & 0.633 & 0.011 & $-0.384$ & 0.007\\
2274.85574 & $-0.859$ & 0.006 & 0.633 & 0.015 & $-0.383$ & 0.008\\
2274.86428 & $-0.863$ & 0.008 & 0.634 & 0.015 & $-0.384$ & 0.008\\
2280.86288 & $-0.849$ & 0.006 & 0.602 & 0.012 & $-0.394$ & 0.007\\
2280.87142 & $-0.834$ & 0.006 & 0.636 & 0.013 & $-0.392$ & 0.007\\
2289.86232 & $-0.795$ & 0.005 & 0.623 & 0.010 & $-0.390$ & 0.006\\
2289.87087 & $-0.783$ & 0.006 & 0.618 & 0.012 & $-0.387$ & 0.008\\
2296.87269 & $-0.750$ & 0.005 & 0.619 & 0.010 & $-0.383$ & 0.006\\
2296.88122 & $-0.741$ & 0.006 & 0.589 & 0.013 & $-0.379$ & 0.008\\
\hline}

\Section{Time Delay}
Looking at the light curves of both images presented in Fig.~1 one can easily 
see that the light curve of image A is much more variable than that of image 
B. It has many features that have no counterparts in the light curve of image 
B. On the other hand, one can also notice a broad 0.3~mag depression in the 
light curve which occurred near ${{\rm HJD}=2452000}$ and is present in both 
light curves. This feature occurred first in the light curve of image B and 
was followed by image A and is the same as that noted by Ofek and Maoz (2003). 
This strongly suggests variability arising from variability of the lensed 
quasar. The well defined shape and long duration of this feature should allow 
determination of time delay. 

Unfortunately, as already mentioned, the feature occurred during 2000/2001 
season of the quasar visibility period when the OGLE project underwent a major 
hardware upgrade. Therefore only infrequent observations of HE1104\,--1805 were 
possible during that time. As a result there are relatively few points 
covering this feature, especially compared to the coverage during the previous 
observing seasons. 

Because the determination of time delay using the entire dataset of images A 
and B would be dominated mainly by the data taken before the feature occurred 
such an approach would repeat unsuccessful attempt of Schechter \etal (2003). 
Therefore, we limited our data sample to 51 points, starting at ${{\rm HJD}= 
2451489.8}$, \ie spanning more than a half of the time of all collected data. 

To find the time delay between images A and B we used standard $\chi^{2}$ 
minimalization. The light curve of image A is steeper than that of image B and 
image A is also brighter than image B. Therefore we followed Ofek and Maoz 
(2003) approach and $\chi^{2}$ fit is described with the following formula: 
$$\chi^2=\sum\frac{(m_{A(t)}-m_{B(t+\tau)}+\alpha(t-t_{\rm mid})+\Delta m)^2}
{\sigma_{A(t)}^2+\sigma_{B(t+\tau)}^2}\eqno(1)$$
where $m_t^A$ and $\sigma_{A(t)}$ are the magnitude of image A and its error, 
respectively, as a function of time $t$; $m_{t+\tau}^B$ and 
$\sigma_{B(t+\tau)}$ are the magnitude and error of image B, but at the moment 
of time shifted by time delay $\tau$; $t_{\rm mid}$ equals to 2451525.378 and is 
defined as HJD midpoint between the first and last observations in the dataset 
analyzed by Ofek and Maoz (2003). It is defined in this manner so that both 
results could be directly compared. The remaining parameters are the linear 
trend of image A, $\alpha$, and magnitude shift between both images, $\Delta 
m$. 

Before the fitting procedure was applied all data points were nightly 
averaged. We decided to make the interpolation of the light curve of image B. 
The measurements of image B  have larger errors because this component is 
fainter, but, on the other hand, its light curve is much smoother than the 
light curve of image A. 

We applied similar interpolation method as used in Schechter \etal (2003): if 
in data taken within 20 days of the desired time we found two points, we 
fitted a straight line; if there were three or more we fitted a parabola and 
in other cases we took a simple average. 

The fit was performed for all three parameters $\tau$, $\alpha$ and $\Delta m$ 
simultaneously and the global minimum was found at the values of ${-
157}$~days, 0.064~mag/yr and 1.555~mag, respectively. The minimum had 
${\chi^2=95.6}$, what yields ${\chi^2/{\it dof}=1.99}$, where {\it dof} is a 
number of degrees of freedom and equals here to 48. Fig.~2 shows the 
$\chi^2/{\it dof}$ as a function of $\tau$ for $\alpha$ and $\Delta m$ fixed 
at best fitting values. 
\begin{figure}[htb]
\centerline{\includegraphics[width=9cm]{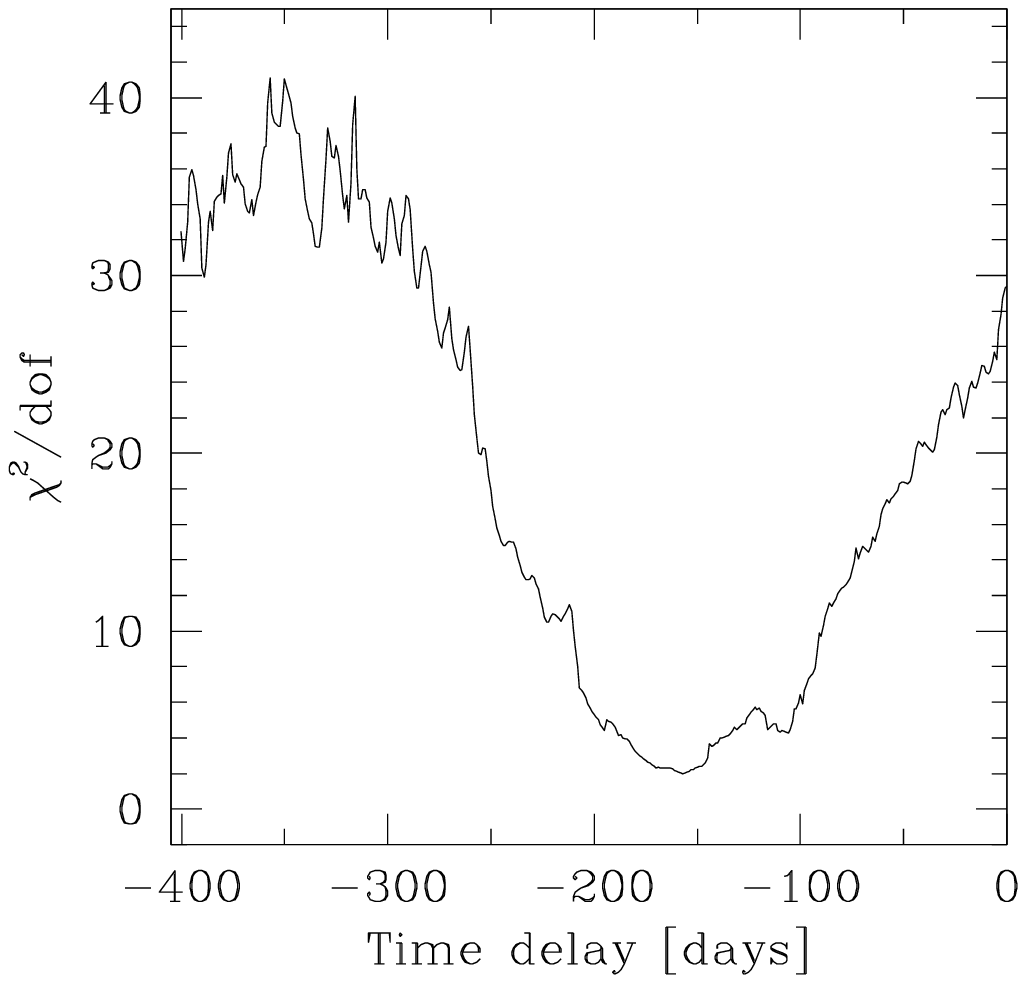}}
\FigCap{$\chi^2/{\it dof}$ as a function of time delay for $\alpha$ and 
$\Delta m$ fixed at the best fitting values (0.064~mag/yr and 1.555~mag, 
respectively). The minimum of ${\chi^2/{\it dof}=1.99}$ is found for ${\tau=-
157}$~days.} 
\end{figure} 

Fig.~3 presents the light curve of image A  with overplotted light curve of 
image B, slope corrected and time and magnitude shifted. Only the subset of 
data that was used for determination of the fit parameters is plotted in 
Fig.~3.
\begin{figure}[htb]
\centerline{\includegraphics[width=13cm]{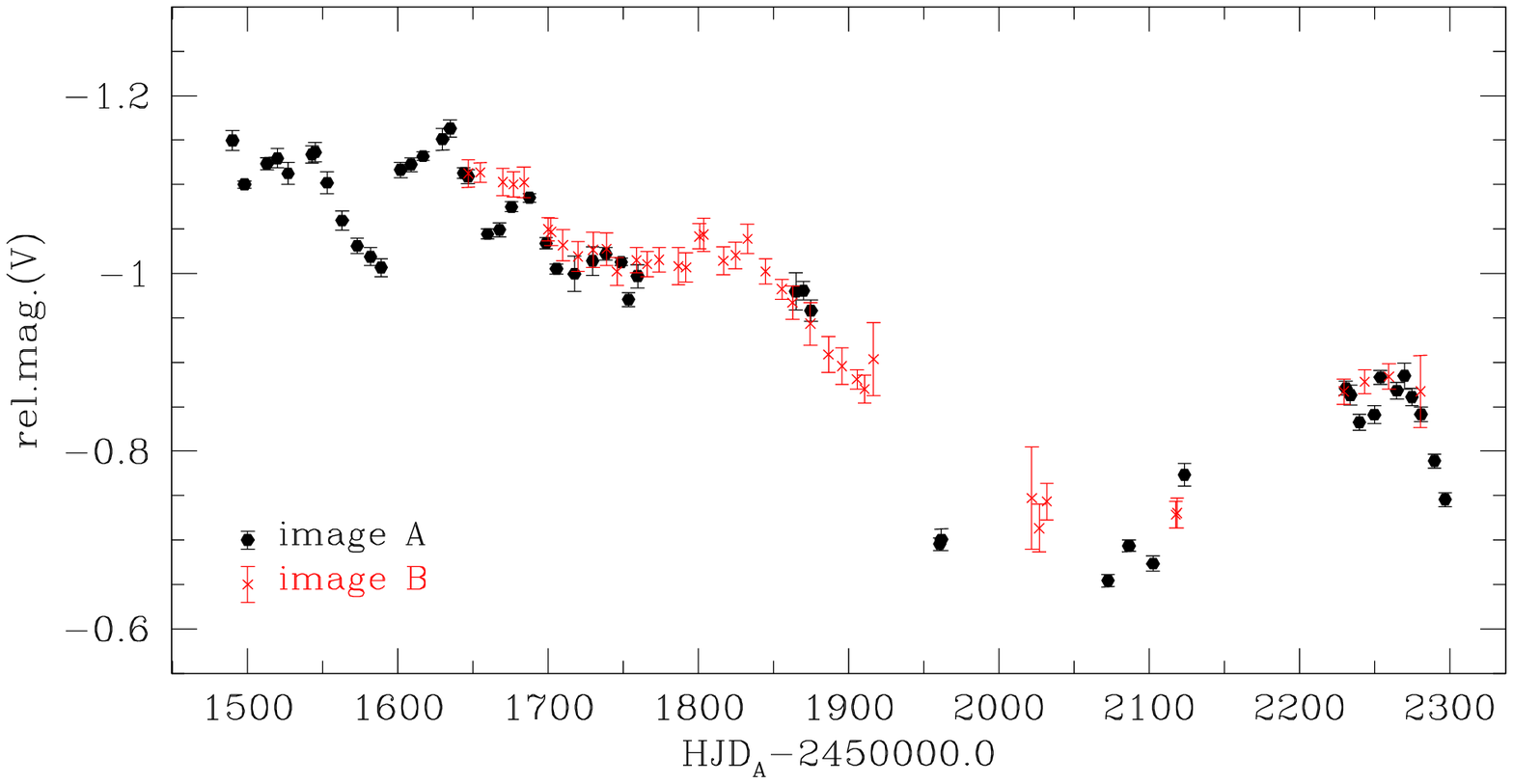}}
\FigCap{Superposition of the light curve of image A with the slope corrected 
(${\alpha=0.064}$~mag/yr), time (${\tau=-157}$~days) and magnitude 
(${\Delta m=1.555}$~mag) shifted light curve of image B. Only data used by 
the fitting procedure are shown.} 
\end{figure} 
\begin{figure}[htb]
\centerline{\includegraphics[width=12.5cm]{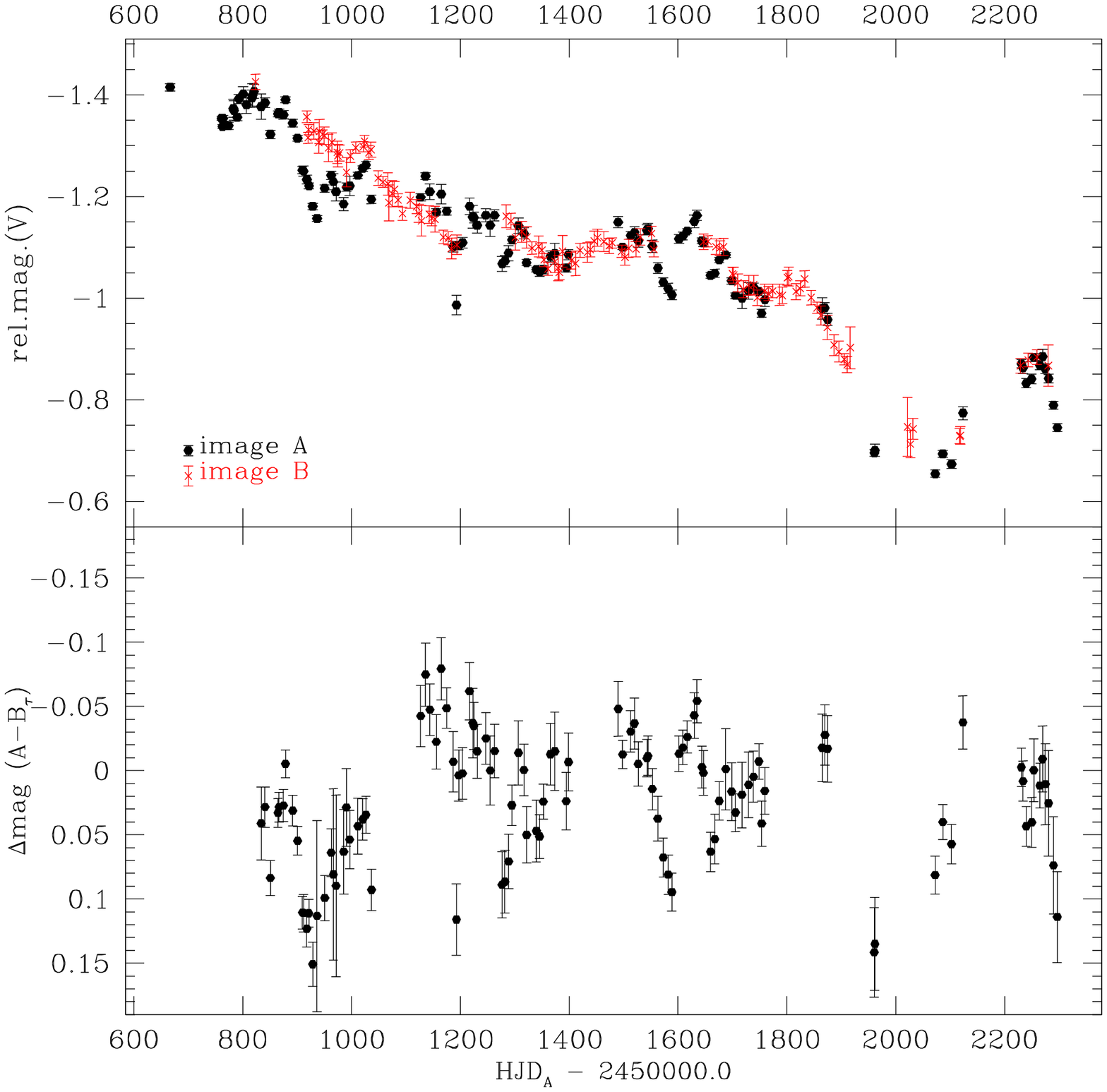}}
\FigCap{Upper panel: as in Fig.~3, but with the best fit parameters 
applied to the entire datasets. Lower panel: the difference between 
light curve of image A and slope corrected and time and magnitude shifted 
light curve of image B.}
\end{figure} 

The upper panel of Fig.~4 presents the composite light curve of both 
images  with the best fit parameters applied for the entire dataset. One can 
notice that in spite of strong microlensing activity in the light curve of 
image A the repeatability of smaller long term features in both light curves 
is rather good what assures that the time delay determination is sound. 

The lower panel of Fig.~4 shows the difference between points of the 
light curve of image A and light curve of image B slope corrected, time and 
magnitude shifted and interpolated to the same HJD. Except for clear 
microlensing scatter in the residuals no long term trend can be noticed. The 
{\it rms} of residuals is equal to 0.086~mag. 

From $\chi^2$ analysis we estimated the error of the time delay value to be 
equal to ${\pm21}$ days. However, this formal error seems to be somewhat 
overestimated when we perform the ``chi-by-eye'' fitting approach. We checked 
possible time delays by overplotting light curves of both images and concluded 
that the error should not be larger than ${\approx10}$~days. Because this 
result can be somehow subjective we assume further the formal error value of 
${\pm21}$~days. 

\Section{Discussion}
The long term monitoring of HE1104\,--1805  gravitational lens during the OGLE 
survey has finally accomplished its original goal. We determined the time delay 
between the light curve of image A and B to be equal to ${-157\pm21}$ days 
where the error is the formal error of $\chi^2$ minimalization. In practice it 
can be somewhat smaller. 

Our result is in excellent agreement with that of Ofek and Maoz (2003) who 
obtained ${\tau=-161^{+7+34}_{-7-11}}$~days (68\% and 95\% confidence level). 
Their result was based on large part of the OGLE dataset (Schechter \etal 
2003) supplemented with {\it R}-band observations covering the parts of the 
light curve crucial for time delay determination (2000--2001 season). One 
should, however, note that lower accuracy of photometry and necessity of 
intercalibration of the OGLE {\it V} and {\it R}-band data might be a source 
of additional uncertainty of the Ofek and Maoz (2003) analysis. 

The entire OGLE dataset provides now homogeneous {\it V}-band coverage of the 
photometric behavior of HE1104\,--1805 between August~1997 and January~2002. 
Thus our time delay determination should be less prone for systematic errors. 

Additional photometry of HE1104\,--1805 presented in this paper nicely confirms 
the features in the light curves noted by Ofek and Maoz (2003). Basically 
identical time delay from our analysis indicates that the time delay of 
HE1104\,--1805 is now determined with good confidence and this gravitational 
lens increases the still small sample of lenses with well determined time 
delays. However, one should note slightly different linear trend between the 
light curves of images A and B. Our value of 0.064~mag/yr is by 0.02~mag 
larger than that of Ofek and Maoz (2003). It is possible that the discrepancy 
is caused by less uniform dataset used by Ofek and Maoz (2003) in their 
analysis. 

\Acknow{The paper was partly supported by the  Polish KBN grants: 2P03D02523 
to {\L}.~Wyrzykowski and 2P03D02124 to A.\ Udalski and by a US NSF grant 
AST-0206010 to P.~Schechter. Partial support for the OGLE project was provided 
by the NSF grants AST-0204908 and NASA  grant NAG5-12212 to B.~Paczy\'nski.}

\end{document}